\documentclass[12pt]{article}
\usepackage{amsmath,amsfonts,amssymb,a4,epsfig,color}

\newcommand{\R}{{\mathbb{R}}}
\newcommand{\C}{{\mathbb{C}}}

\def\ha{\frac{1}{2}}

\def\pa{\partial}
\def\ra{\rightarrow}
\def\preuve{\begin{proof}} 
\def\ga{\alpha}
\def\gb{\beta}

\def\gl{\lambda}
\def\go{\omega}

\def\OPD{~$\Psi DO$~}

\newtheorem{lemm}{Lemma}
\newtheorem{prop}{Proposition}
\newtheorem{rem}{Remark}
\newtheorem{coro}{Corollary}
\newtheorem{theo}{Theorem}

\newenvironment{demo}{\noindent {\it Proof.--}
      \begin{quotation}\noindent}{\end{quotation}\hfill$\square $}

\begin{document}

\title{Semi-classical trace formulas and 
heat expansions}

\author{Yves Colin de Verdi\`ere \footnote{Institut Fourier,
 Unit{\'e} mixte
 de recherche CNRS-UJF 5582,
 BP 74, 38402-Saint Martin d'H\`eres Cedex (France);
yves.colin-de-verdiere@ujf-grenoble.fr
}}


\maketitle

\section*{Introduction}
There is a strong similarity between the expansions of the heat
kernel
as worked out by people in Riemannian geometry in the seventies
(starting with the famous ``Can one hear the shape of a drum''
by Mark Kac \cite{Kac},
 the Berger paper  \cite{Berger} and the Mc-Kean-Singer paper \cite{McK-Si})
and the so-called semi-classical trace formulas developed by people
in semi-classical analysis (starting with Helffer-Robert \cite{He-Ro}).
In fact, this is  not only a similarity, but, as we will prove, each of these
expansions, even if they differ when expressed numerically
for  some example,    can be deduced from the other one
as formal expressions of 
the fields.

Let us look first at the {\it heat expansion} on a smooth closed Riemannian
manifold
of dimension $d$,
$(X,g)$, with the  (negative) Laplacian $\Delta_g$\footnote{In this
  note,
we will not follow the usual sign convention of geometers, but the
convention
of analysts}.
The heat kernel $e(t,x,y)$, with  $t>0$
and $x,y \in X$,
is  the Schwartz kernel of 
${\rm exp}(t \Delta_g) $:
the solution of the heat equation
$u_t - \Delta _g u =0$ with initial datum  $u_0$ is given
by 
\[ u(t,x)=\int_X e(t,x,y) u_0(y) |dy|_g ~.\]
The function 
 $e(t,x,x) $ admits, as $t\ra 0^+$, the following
asymptotic expansion:
\[ e(t,x,x)\sim (4 \pi t)^{-d/2}\left(
1 + a_1(x)t+\cdots +  a_l (x)t^l + \cdots \right) ~.\]
The  $a_l$'s are given explicitly in \cite{Gilkey2}, page 201,
for $l\leq 3$ and are known for $l\leq 5$ \cite{Avramidi,vdV}.
See also the related works by Hitrik and  Polterovich 
\cite{Hi,H-P1,H-P2,Po}. 
They are  universal polynomials in the components of the
curvature
tensor and its  co-variant derivatives.
For example $a_0 =1 $,
$a_1= \tau_g/6  $ where $\tau_g $ is the scalar curvature.

The previous asymptotic expansion gives the expansion
of the trace by integration over $X$ and has been used as an important
tool in spectral geometry:
\[ {\rm trace} (e^{t\Delta_g})=\int_X e(t,x,x)|dx|_g= 
\sum_{k=1}^\infty  e^{\gl_k t}~,\]
where  $-\gl_1=0 \leq -\gl_2 \leq \cdots \leq -\gl_k \leq \cdots $
is the sequence of eigenvalues of $-\Delta _g$ with the usual
convention
about multiplicities. 
If $d=2$, this gives
 \[ {\rm trace} (e^{t\Delta_g})=\frac{1}{4\pi t}\left({\rm Area}(X)
+\frac{2\pi \chi(X)}{6}t+0(t^2) \right)~,\]
where $\chi(X)$ is the Euler characteristic of $X$.

There is an extension of the previous expansion in the case of Laplace
type operators on fiber bundles: the coefficients of the
expansion are then polynomials in the co-variant derivatives
of the curvature of the metric and  of the connection on the
fiber bundle.
The heat  expansion  can be re-interpreted  as 
an expansion of the Schwartz kernel of $f(-\hbar^2 \Delta _g )$
on the diagonal $x=y$
in powers of $\hbar$ with $f(u)={\rm exp}(-u)$ and
$t=\hbar ^2$. This is a particular case of the
semi-classical trace.

Let us describe the semi-classical setting  in the  flat
case:
$\widehat{H_\hbar} $ is a self-adjoint $\hbar-$pseudo-differential operator
 with Weyl symbol
$H(x,\xi) $ in some open domain $X$  in $\R^d$, or more generally 
on a Riemannian manifold. 
Let $f\in {\cal S} (\R )$ and look at
$f(\widehat{H_\hbar} )$. Under some suitable assumptions
(ellipticity at infinity in $\xi$)
on $H $, $f(\widehat{H_\hbar}) $ is a pseudo-differential operator
whose Weyl symbol $f^\star (H)$ is a  formal power series in 
$\hbar $,  given, using the Moyal product denoted by $\star$,  by the
following formula (see \cite{Gracia} for explicit formulas
and Section 4.2 for a proof, see also \cite{Charles})
 at the point $z_0 \in T^\star X$:
\begin{equation}\label{equ:gracia}
 f^\star (H)(z_0)=(2\pi \hbar)^{-d}
\left( \sum_{l=0}^\infty \frac{1}{l!}f^{(l)}(H(z_0)) \left( H-H(z_0) \right)
 ^{\star  ~ l}(z_0)\right)~.\end{equation}
From the previous formula, we see that the symbol
of  $f(\widehat{H_\hbar}) $  at the point $z$ depends only of the
Taylor expansions of $H$ at the point $z$ and
 of $f$ at the point $H(z)$. 
In the paper \cite{H-P}, the authors study the case of the magnetic
Schr\"odinger operator whose Weyl symbol is $H_{a,V}(x,\xi)=
\sum _{j=1}^d (\xi_j
-a_j(x))^2+V(x)$
and show that the Schwartz kernel
of  $f(\widehat{H_{\hbar, a,V}}) $  at the point $(x,x)$ admits an
asymptotic  expansion of the form
\[ [f(\widehat{H_{\hbar, a,V}})](x,x)=(2\pi \hbar )^{-d}
\left[ \sum_{j=0}^\infty \hbar^{2j} 
\left( \sum _{l=0}^{k_j}\int_{\R^d }
  f^{(l)}( \| \xi \| ^2 +V(x)) Q_{j,l}^{a,V}(x,\xi)|d\xi|
 \right) \right] \]
where the $Q_{j,l}^{a,V}(x,\xi)$'s are 
polynomials in $\xi $ calculated from
 the Taylor expansions of the magnetic field  $B=da$
and $V$ at the point $x$.  
The proof in \cite{H-P}
 uses a pseudo-differential calculus adapted to the magnetic
field.

We will  give a simplified version of the expansion replacing the
(non unique) $Q_{j,l}^{a,V}(x,\xi) $'s by functions $P_{j,l}^{B,V}(x) $'s
which are uniquely defined and are given by universal
$O(d)$-invariant
polynomials of the Taylor expansions of $B$ and 
$V$ at the point $x$.
We present then   two  ways to compute the $P_{j,l}^{B,V}$'s:
\begin{itemize}
\item 
we can first use Weyl's invariant theory (see Gilkey's book
\cite{Gilkey2}) in order to reduce the
problem to the determination of a finite number of numerical
coefficients; then simple examples, like harmonic oscillator
and constant magnetic field, allow to determine (part of) these
coefficients.
\item The $P_{j,l}^{B,V}$'s are related in a very simple way to 
the coefficients of the heat expansion; it is possible to compute
the $P_{j,l}^{B,V}$'s from the knowledge of the $a_l$'s for
$j+1 \leq l \leq 3j$. This is enough to re-compute
the coefficient of $\hbar^2$ and also,  in principle,
  the coefficients of $\hbar ^4 $ in the
expansion, because the $a_l$'s are known up to $l=6$ in the
case of a flat metric (see \cite{vdV}).
\end{itemize} 

In this note, we will first describe precisely the semi-classical
expansion for Schr\"odinger operators (in the case 
of an Euclidian metric) and the properties
of the functions $P_{j,l}^{B,V}(x)$'s. Then, we will show how 
to compute the $P_{j,l}^{B,V}(x)$'s using an adaptation of the method
used
for the heat kernel (Weyl's  Theorem on invariants and explicit
examples). Finally,  we will explain how the $a_l$'s are related
to the $P_{j,l}^{B,V}(x)$'s. This gives us two proofs of the
main formula given by Helffer and Purice in \cite{H-P}; this paper 
 was the initial motivation to this work. 

\subsection*{Acknowledgments}
{\it I thank the referee for his careful reading of the paper which
forced me to make the statements and proofs  more precise. Many thanks also
to Johannes Sj\"ostrand for his help in clarifying some points
of semi-classical analysis and for allowing me to present them in 
 an 
Appendix to this paper.}

\section{Semi-classical trace for Schr\"odinger operators}
In what follows, $X$ is an open domain in $\R^d$, equipped with the
canonical
Euclidian metric, 
and $\Omega ^k(X)$ will denote the space of smooth exterior differential forms
in $X$.
Let us give a Schr\"odinger operator, with a smooth magnetic field
$B=\sum_{1\leq i<j \leq d} b_{ij}dx_i\wedge dx_j$
 (a closed real 2-form)  and a smooth electric potential $V$
(a real valued smooth function) in $X$. We assume that $V$ is bounded
from below.
 We will assume also that  the 2-form $B$ is exact and  can be written
 $B=da $ and we introduce the Schr\"odinger operator defined by  
\[ {H_{\hbar,a,V}}=
 \sum_{j=1}^d \left( \frac{\hbar}{i}\frac{ \pa }{\pa x_j }
-a_j(x) \right)^2 + V(x)  ~. \]
The Weyl symbol of ${H_{\hbar,a,V}} $
is $H_{a,V}(x,\xi)=\| \xi -a (x) \|^2 +V(x)$.
We denote by  $\widehat{H_{\hbar,a,V}}$ a self-adjoint extension of
 ${H_{\hbar,a,V}}$ in $L^2(X,|dx|)$. 
Let us give $f \in {\cal S}(\R )$
and $\phi \in C_o^\infty (X)$  and consider
the 
trace of $\phi f( \widehat{ H_{\hbar,a,V}})$ as a distribution on $X \times \R $
(the density of states):
\[ {\rm Trace}(\phi f( \widehat{ H_{\hbar,a,V}}))=
\int_X Z_{\hbar,a,V}(g)(x)\phi(x) |dx| ~,\]
where  $Z_{\hbar,a,V}(g)(x)$ is the value at the point $(x,x)$
of the Schwartz kernel of $ f( \widehat{ H_{\hbar,a,V}})$. 
\begin{theo}\label{theo:main}
We have the  following asymptotic expansion in power of $\hbar$:
\[ Z_{\hbar,a,V} ( g)(x) \sim
(2\pi \hbar)^{-d} \left[\int _{\R^d} f ( \| \xi \| ^2 +V(x)) |d\xi |
 +  \sum _{j=1}^\infty \hbar^{2j}
\left( \sum _{l=j+1}^{l=3j}P_{j,l}^{B,V} (x) 
 \int _{\R^d} f^{(l)} ( \| \xi \| ^2 +V(x)) |d\xi |\right)
\right]~.\]
We have 
 the explicit formulas
\[ P_{1,2}^{B,V}=-\frac{1}{6}\left( \Delta V + \| B \| ^2 \right) ~,\]
\[ P_{1,3}^{B,V}=-\frac{1}{12}\| \nabla V \|^2 ~,\]
\[ P_{2,3}^{B,V}=-\frac{1}{180}\left(
8 \| \nabla B \|^2 + \| d^\star B \| ^2 + 12\langle \Delta B|B \rangle
+ 3 \Delta ^2 V \right)~.\] 
Here $ \| B \| ^2 = \sum _{1\leq i< j \leq d } b_{ij}^2 $, $d^\star:
\Omega ^2(X) \ra \Omega ^1(X) $
is the formal adjoint of $d$ used in the definition of the Hodge
Laplacian
on exterior forms.  
If $d=3$, $ \| B \|$ is the Euclidean norm of the vector field
associated to $B$.

The $P_{j,l}^{B,V} (x)$ are polynomials of 
the derivatives of $B$ and $V$ at the point $x$.
Moreover, if $\lambda, ~\mu,~ c $
are constants and  we define  $\lambda^\star( f)(x)=f(\lambda x )$,
we have the following scaling properties:
\begin{enumerate}
\item 
$P_{j,l}^{\lambda .  \lambda^\star( B), \lambda^\star( V)}(x)=
 \lambda ^{2j}P_{j,l}^{B, V}(\lambda x)$. This will be used
with $x=0$.
\item 
$ P_{j,l}^{\mu B,\mu^2 V}(x)= \mu ^{2(l-j)}P_{j,l}^{B, V}(x)$
\item  $P_{j,l}^{B,V+c}(x)=P_{j,l}^{B,V}(x)$
\item  $P_{j,l}^{- B, V}(x)=P_{j,l}^{B, V}(x)$.
\item The $P_{j,l}^{B,V}$'s are
invariant by the natural action of the orthogonal group $O(d)$ on the Taylor
expansions of $B$ and $V$ at the point $x$.
\end{enumerate}
\end{theo}
\begin{rem} From the statement of the previous Theorem, we see
that the expansion of the density of states is independent of the
chosen
self-adjoint extension.
\end{rem}

As a consequence, we can get the following full trace expansion
under some more assumptions:
\begin{coro} Let us assume that $E_0=\inf V < E_\infty
=\liminf_{x\ra \pa X}  V(x)$ and that we have chosen the Dirichlet
boundary
conditions.
Let $f\in C_o^\infty (]-\infty, E_\infty [)$,
then the trace  of $f(\widehat{H_{\hbar,a,V}})$
admits the following asymptotic expansion
 \[ {\rm Trace}(f(\widehat{H_{\hbar,a,V}})) \sim
(2\pi \hbar)^{-d}\int_X \left( \int _{\R^d} f(\| \xi \|^2 +V(x))|d\xi|
  + \cdots  \right. \]
 \[\left.  \cdots  \sum _{j=1}^\infty \hbar^{2j} \\
\sum _{l=j+1}^{l=3j}P_{j,l}^{B,V} (x) 
 \int _{\R^d} f^{(l)} (\| \xi \|^2 +V(x)) |d\xi |
\right)|dx|~,\]
The coefficient of $\hbar^2$
can be written as  
\[ -\frac{1}{12}\int_{X\times \R^d} 
 f^{(2)} (\| \xi \| ^2 +V(x))\left( \Delta V(x) +
2 \| B (x)\| ^2 \right) |dxd\xi | ~.\]
\end{coro}
The expansion follows from \cite{He-Ro}.
An integration by part in $x$
gives
\[ \int _X f^{(3)}(\| \xi \| ^2 +V(x) ) \| \nabla V (x)\|^2 |dx|
=- \int _X f^{(2)}(\| \xi \| ^2 +V(x) ) \Delta V (x) |dx|~.\]

\section{Existence of the $\hbar-$expansion of
$Z_{\hbar,a,V}$}

Using Theorem \ref{theo:js} in the Appendix,
 we can work in $\R^d$ with $a$ and $V$
compactly supported.
The existence of the expansion is known in general
from \cite{He-Ro} and the calculus of the symbol of
$f(\widehat{H_{\hbar,a,V}})$.
We get 
\[ \int _X Z_{\hbar,a,V}( f)(x)\phi(x) |dx|=
(2\pi \hbar )^{-d} \left(
\sum_{j=0}^\infty  \hbar^{2j} \sum _{l=0}^{k_j}
\int \phi(x) f^{(l)}(H_{a,V}(x,\xi))Q_{j,l}(x,\xi) |dx d\xi | \right)\]
where the $Q_{j,l}(x,\xi)$'s are polynomials in the Taylor
expansion of $H_{a,V}$ at the point $(x,\xi)$.
The previous expansion is valid for any (admissible) pseudo-differential
operator.
In the case of Schr\"odinger operators we can make  integrations
by part in the integrals
$ \int  f^{(l)}(H_{a,V}(x,\xi))Q_{j,l}(x,\xi) |d\xi| $
which reduces to a similar formula where we can replace
the $Q_{j,l}(x,\xi) $'s by  $P_{j,l}(x) $'s.
This is based on the expansion of $Q_{j,l}$ as a polynomial in $\xi$
in powers of $(\xi-a)$: odd powers give $0$  and 
even powers can be reduced  using 
\[ d_\xi \left( (\xi_j  -a_j )f^{(l)}(H_{a,V})\iota(\pa _{\xi_j})d\xi \right) =
2 \| \xi_j -a_j \|^2 f^{(l+1)}(H_{a,V})d\xi + f^{(l)}(H_{a,V})d\xi~.\]

We have only to check that the powers of $\xi$ in $Q_{j,l}(x,\xi)$ are
less
than $l$: this is based on Equation (\ref{equ:gracia}).
The coefficients of the  $l-$th Moyal power  of $H_{a,V}(z)-H_{a,V}(z_0)$
are homogeneous polynomials  of degree $l$ in the derivatives
  of $H_{a,V}(z)$. At the point $z=z_0$ only
derivatives of order $\geq 1$ are involved.
They are all of degree  $\leq 1$ in $\xi$.
Using Gauge invariance at the point $x$
(Section \ref{sec:gauge}), we can assume that $a(x)=0$.

\section{Gauge invariance}\label{sec:gauge}

If $S:X\ra \R $ is a smooth function,
we have
\[ {\rm Trace}(\phi e^{-iS(x)/\hbar}f(\widehat{H_{\hbar,a,V}})
e^{iS(x)/\hbar})= {\rm Trace}(\phi f(\widehat{H_{\hbar,a,V}}))~\]
and
\[ e^{-iS(x)/\hbar}f(\widehat{H_{\hbar,a,V}})
e^{iS(x)/\hbar}=f( \widehat{H_{a+dS ,V}})~.\]

Hence, we can chose any local gauge $a$ in order to compute the
expansion: using the synchronous gauge (see Section \ref{sec:sync}),
we get the individual terms
\[ \int f^{(l)}(H_{0,V}) P_{j,l}^{B,V}(x) |d\xi | ~\]
for the expansion, where the  $P_{j,l}^{B,V}(x)$
depend only of the Taylor expansions of $B$ and $V$
at the point $x$.

\section{ The synchronous gauge}\label{sec:sync}


The main idea is to find an appropriate gauge $a$ adapted to the
point $x_0$ where we want to make the symbolic computation.
In a geometric language,
 we use the trivialization of the bundle by parallel transportation
along the rays: the potential $a$ vanishes on the radial vector field
\footnote{This gauge is sometimes called the {\rm Fock-Schwinger} gauge;
in \cite{A-B-P}, it is called the  {\rm synchronous framing}  }.
Here, this is simply the fact that, for any closed 2-form $B$ on $\R^2$,
there exists an  unique 1-form $a=\sum_{j=1}^d a_j dx_j$ so that $da=B$
and $\sum_{j=1}^d x_j a_j =0$.

We will do that for the Taylor expansions degree by degree.
In what follows we will use a decomposition for 1-forms, but it works
also for $k-$forms.

Let us denote by $\Omega ^k_N$ the finite dimensional vector
space of $k$-differential forms on $\R^d$ whose coefficients
are homogeneous polynomials of degree $N$ and by $W=\sum_{j=1}^d x_j 
\frac{\pa }{\pa x_j }$ the radial vector field.
The exterior differential induces a linear
map from $\Omega ^k_N $ into $\Omega ^{k+1}_{N-1}$
and the inner product $\iota (W)$ a map
from $\Omega ^{k}_{N}$ into  $\Omega ^{k-1}_{N+1} $.
They define  complexes which are exact 
except at $k=N=0$. Moreover, we have a situation similar to Hodge
theory:   
\[ \Omega ^k_N= d\Omega ^{k-1}_{N+1}\oplus 
\iota(W) \Omega^{k+1}_{N-1} ~.\]
This is due to  Cartan's formula:
the Lie derivative of a form $\omega  \in \Omega _N^{k}$
satisfies, from the direct calculation,
${\cal L}_W \go = (k+N)\go $,
and, by Cartan's formula, 
${\cal L}_W \go= d(\iota (W) \go )+ \iota (W) d\go $.
So 
\[ \go =\frac{1}{k+N} \left(  d(\iota (W) \go )+ \iota (W) d\go
\right)~.\]
It remains to show that this is a direct sum:
if $\go = d\ga =\iota (W) \gamma $,
we have $\iota (W)\go=0$ and $d\go=0$; from the previous decomposition,
we see that $\go=0$.
Let us denote by $J^N\go$, where  $\go$ is a differential form of degree $k$,
the form in $\Omega _N^k$ which appears in the Taylor expansion
of $\omega $.   

We get
\begin{prop} If $P(J^0a ,J^1a, \cdots , J^Na)$
 is a polynomial in the Taylor expansion
of the 1-form $a$ at some order $N$ which is invariant by $a \ra a+ dS
$,
$P$ is independent of $J^0a$ and
 \[ P\left(J^1a, \cdots , J^Na\right)=P\left( \ha J^1 \iota (W) B,
 \cdots,\frac{1}{N+1} J^N \iota (W) B\right) \]
is a polynomial of the Taylor expansion of $B$ to the order $N-1$.

\end{prop}

\section{Properties of the $P_{j,l}$'s}
\subsection{Range of $l$ for $j$ fixed}

From the scaling properties, we deduce
that, in  a monomial  
\[ D^{\ga_1} B_{i_1,j_1}\cdots D^{\ga_k} B_{i_k,j_k} 
  D^{\gb_1} V \cdots
D^{\gb_m} V~,\]
belonging to  $P_{j,l}$, we have 
$k+2m=2(l-j)$ 
and
$ k +|\ga_1|+\cdots +|\ga_k|+ |\gb_1|+\cdots +|\gb_m|    = 2j $.
Moreover, for $ j\geq 1$,  $k+m \geq 1$ and  $|\gb_p|\geq 1$.
Hence  $j+1\leq l \leq 3j $.
The previous bounds are  sharp:
take the monomials $\Delta ^{j}V$ and $\| \nabla V \|^{2j}$
which give $l=j+1$ and $l=3j$.

\subsection{Invariance properties}

\begin{enumerate}
\item
Let us assume that we look at the point 
$x=0$ and consider the operator
$D_\mu (f)(x)=f(\mu x)$.
We have 
\[ D_\mu \circ \hat{H}_{\hbar, A,V}\circ D_{1/\mu}=
 \hat{H}_{\hbar/\mu, A\circ D_\mu,V\circ D_\mu}~.\]
The same relation is true for any function $f(\hat{H}_{\hbar, A,V})$
and then we have, looking at the Schwartz kernels
and using the Jacobian $\mu^d$ of $D_\mu$:
\[ P_{j,l}^{B,V}(0) \int _{\R^d} f^{(l)}(\| \xi \|^2 +V(0))|d\xi| =
 \mu ^{-2j} P_{j,l}^{\mu. \mu^\star B,\mu^\star V}(0)
 \int _{\R^d} f^{(l)}(\| \xi \|^2 +V(0))|d\xi|~.\]
\item 
We have
\[ \widehat{H}_{\hbar,\mu a , \mu^2 V}=
\mu ^2  \widehat{H}_{\frac{\hbar}{\mu}, a ,  V}~.\]
\item Changing $V$ into $V+c$ gives a translation by $c$ in the
  function $f$ but does not change the $P_{j,l}^{B,V}$'s. 
\item Changing $B$ into $-B$ gives a complex conjugation in
the computations. The final result is real valued.
\item
Orthogonal invariance is clear:  an orthogonal change of
coordinates
around the point $x$ preserves the density of states.
\end{enumerate}
\subsection{The case $d=2$}

We deduce from the scaling properties and invariance
by the orthogonal group, that there exists
constants $a_d,~ b_d,~ c_d$ so that
$P_{1,2}^{B,V}(x)= a_d \Delta V + b_d \| B\| ^2 ,~
P_{1,3}(x)=c_d \| \nabla V \| ^2 $.

\section{Explicit examples}
The calculation for the harmonic oscillators and the constant
magnetic fields allows to determine the constants   $a_d,~ b_d,~ c_d$.

\subsection{Harmonic oscillators}
Let us consider $\Omega = -\hbar^2 \frac{d^2}{dx^2}+ x^2 $ with $d=1$.
The kernel of $P(t,x,y)$ of $ {\rm exp}(-t\Omega )$ is given by the
Mehler formula:
\[ P(t,x,y)=(2 \pi \hbar \sinh (2t\hbar ))^{-\ha}
{\rm exp}\left(-\frac{1}{2\hbar \sinh (2t\hbar)}\left( \cosh (2t \hbar)(x^2
  +y^2)-2xy \right)\right)~.\] 
Hence
\[ P(t,x,x)\sim (2\pi \hbar)^{-1}e^{-tx^2} \left( \int_\R e^{-t\xi^2}
  d\xi \right)
\left( 1- \hbar^2(t^2    -t^3 x^2 )/3 + 0(\hbar^4) \right)~.\]
Hence
$P_{1,2}(x)=-V''(x)/6,~P_{1,3}(x)= -V'(x)^2 /12,  $. 

Similarly, in dimension $d>1$, we get
$P_{1,2}(x)=-\Delta V(x)/6,~P_{1,3}(x)= -\| \nabla V \|^2 /12,  $. 

\subsection{Constant magnetic field}
Let us consider the case of a constant magnetic field $B$
in the plane and denote by $Q(t,x,y)$ the kernel of 
${\rm exp}(-t H_{B,0})$.
We have (see \cite{A-H-S})
\[ Q(t,x,x)= \frac{B}{4 \pi \hbar  \sinh Bt \hbar }~.\] 
Hence the asymptotic expansion
\[  Q(t,x,x)=(2\pi \hbar)^{-2}  \int {\rm exp }(-t\|\xi \|^2) |d\xi|
\left( 1 - t^2 \hbar^2 B^2 /6 +0(\hbar^4) \right) ~,\]
hence
$P_{1,2}(x)=-B^2/6,~ P_{1,3}(x)=0$. 

Using the normal form
$B=b_{12}dx_1 \wedge dx_2 +b_{34}dx_3\wedge dx_4 + \cdots $,
we get in dimension $d>2$,
$P_{1,2}(x)=-\|B \|^2/6,~ P_{1,3}(x)=0$.

\section{Heat expansion from the semi-classical expansions}

We have
$t\widehat{H_{1,a,V}}=\widehat{H_{\sqrt{t},\sqrt{t}a,tV}}$.
Using the expansion of Theorem \ref{theo:main}
with $f(E)=e^{-E}$, we get
easily the point-wise expansion of the heat kernel on the diagonal
as $t\ra 0^+$:
\[ [{\rm exp }(-t\widehat{H_{1,a,V}})](x,x)\sim 
\frac{1}{(4\pi t)^{d/2}}e^{-tV(x)} \left[ \sum_{l=0}^\infty 
\left( \sum _{l/3\leq j \leq l-1} P_{j,l}^{B,V} (x)\right)(-t)^l \right] ~.\]
In particular,  $a_1(x)=-V(x)$ and 
the coefficient $a_2(x)$ is given
by
\[ a_2(x)= \ha V(x)^2 -\frac{1}{6}\Delta V(x) -\frac{1}{6}\| B(x) \|^2
~.\]
This formula agrees with Equation (3) of Theorem 3.3.1 in
\cite{Gilkey2}.

 This gives another way to compute the $P_{j,l}$'s:
if, as power series in $t$,  
\[ \sum_{l=0}^\infty (-1)^l b_l(x)t^l = e^{tV(x)}\left( \sum _{l=0}^\infty
  a_l(x)t^l
\right)~, \]
we have
\[ \sum  _{l/3\leq j \leq l-1} P_{j,l}^{B,V} (x)= b_l(x) ~.\]
$ P_{j,l}^{B,V} $ is the sum of monomials homogeneous 
of degree $2(l-j)$ in $b_l$ where $B$ and its derivatives have
weights $1$ while $V$ and its derivatives have weights $2$.

The heat coefficients $a_l$ on flat spaces are known 
for $l\leq 6$ from \cite{vdV}.
This is enough to check  the term in $\hbar^2 $ (uses $a_2$
and $a_3$) in \cite{H-P}  and
to compute the term in $\hbar^4$ in the semi-classical expansion
(uses the $a_l$'s for $3\leq l \leq 6 $). 

We have also a mixed expansion
writing $t\widehat{H}_{\hbar,a, V}=
\widehat{H}_{\sqrt{t}\hbar,\sqrt{t}a,t V}$,
we get a power series expansion in powers of $\hbar$ and $t$
valid in the domain $\hbar^2 t \ra 0$ and $0< t\leq t_0 $
for the point-wise trace of ${\rm exp}(-t\widehat{H}_{\hbar,a, V})$:
\[ Z_{t,\hbar}(x) \sim \frac{1}{(4 \pi t)^{d/2}}e^{-tV(x)}
\left(1+ 
 \sum _{j\geq 1,~j+1 \leq l \leq 3j} \hbar^{2j}(-t)^l P_{j,l}^{B,V}(x)
\right) ~.\] 
This shows that the integrals
$\int_X V(x)^k |dx|$ and  $\int_X  P_{j,l}^{B,V}(x)|dx|$
are recoverable from the semi-classical spectrum.

\section*{Appendix: functional calculus in domains and self-adjoint
extensions (after  Johannes Sj\"ostrand)}

{\it The content of this Appendix is due to Johannes Sj\"ostrand. I
  thank
him very much for this contribution.}

Let $X\subset \R^d$ be an open set, we say that a linear operator $A$ is a
\OPD ~ in $X$, with Weyl symbol $a$,
 if, for any compact $K\subset X$, $A$ acts on functions
supported in $K$ as a \OPD of Weyl symbol $a$.
\begin{theo} \label{theo:js}
 Let $H_{\hbar,a,V}$ be a Schr\"odinger operator with
  magnetic field given by 
\[ H_{\hbar,a,V}=
 \sum_{j=1}^d \left( \frac{\hbar}{i}\frac{ \pa }{\pa x_j }
-a_j(x) \right)^2 + V(x)  ~, \]
defined in some open domain $X \subset \R^d$. We assume that $a$ and
$V$
are smooth in $X$ and that  $V$ is
bounded from below,  so that  $H_{\hbar,a,V}$
admits some self-adjoint extensions on the Hilbert space
$L^2(X, |dx|)$. One of them will be denoted by
 $\widehat{H_{\hbar,a,V}}$.
Then, for any $f \in {\cal S} (\R)$, $f(\widehat{H_{\hbar,a,V}})$,
given by the functional calculus, is a semi-classical \OPD ~ 
in $X$ whose symbol is given by Equation (\ref{equ:gracia}) and is 
independent of the chosen extension.
\end{theo}

The proof uses a multi-commutator method already used by Helffer and
Sj\"ostrand.
\begin{demo}
We introduce, for $s\in \R$,
 the semi-classical ($\hbar$-dependent) Sobolev spaces
\[ {\cal H}^s_\hbar :=\{ u \in {\cal S}'(\R^d)~|~\|{\rm Op}_\hbar (1+\| \xi
\|^2)^{s/2}u \|_{L^2}<\infty ~\}~\]
with the norm
\[ \|u \|_s:=\| {\rm Op}_\hbar (1+\| \xi
\|^2)^{s/2}u \|_{L^2}~.\]
The ($\hbar$-dependent) norm $\| A \| _{s_1, s_2}$
is the norm of $A$ as linear operator from  $ {\cal H}^{s_1}_\hbar $
to  $ {\cal H}^{s_2}_\hbar $.
A linear operator $K$ is smoothing if, for all $s_1,~s_2$,
$\| K \|_{s_1,s_2}=O(\hbar ^\infty )~.$
This implies that the Schwartz kernel of $K$  is smooth with  all derivatives
 locally $O(\hbar ^\infty )$.
We have the 
\begin{lemm} \label{lemm:sj} Let $Y$ be an open set in $\R^d$.
Let  $P_j=P_j(\hbar),~j=0,1$ be two self-adjoint operators on Hilbert spaces
${\cal H}_j=L^2(X_j,|dx|)$
with $Y\subset \subset  X_0 \subseteq X_1\subseteq \R^d$ 
and 
 with domains ${\cal D}_j$ so that
$C_o^\infty (Y)\subset {\cal D}_j\subset  {\cal H}_j$.
 Let us assume that,  on $C_o^\infty (Y)$, $P_0=P_1=
H_{\hbar, a,V}(=P)$.

Then, for any $f\in C_o^\infty (\R)$,
$ f(P_0)-f(P_1)$ is smoothing on $Y$. In particular, the densities of
states $[f(P_j)](x,x),~j=0,1,$ co\"incide in $Y$ modulo $O(\hbar^\infty)$.
\end{lemm}
Assuming  Lemma \ref{lemm:sj},  Theorem \ref{theo:js}
 follows by extending $a$ and $V$ 
smoothly outside $Y$ so that they have compact support in $\R^d$.
We take  $Y \subset \subset X =X_0 \subset \R^d =X_1$.
It follows that $P_1$ is  essentially self-adjoint and  
the functional calculus for $P_1$  follows then easily from \cite{He-Ro}.
The result is valid  even for $f\in {\cal S}(\R)$ because
 $C_o^\infty $ is dense in  ${\cal S}$ and the 
result of \cite{He-Ro} is valid for $f\in {\cal S}$
and the resulting formulas for the symbols 
 are  continuous w.r. to the topology of  ${\cal
  S}$.
\end{demo}

 {\it Proof of Lemma \ref{lemm:sj}.--}
If $\chi \in C_o^\infty (Y)$, then, for $z\notin \R$
and $j,k\in \{ 0,1 \}$, we have on $L^2(Y)$:
\begin{equation}\label{equ:1}
(P_j-z)^{-1}\circ \chi =\chi \circ (P_k-z)^{-1}
-(P_j-z)^{-1}[P,\chi ](P_k-z)^{-1}~\end{equation}
Let $ \chi _0 \leq \chi_1 \leq \cdots \leq \chi_N $ with, for
$l=0,\cdots, N$, 
$\chi_l\in C_o^\infty (Y) $ and, for
$l=0,\cdots, N-1$, 
$\chi _l(1-  \chi_{l+1})\equiv 0 $.
By iterating Equation (\ref{equ:1})
and using $\chi_{l+1} [P,\chi_{l}]= [P,\chi_{l}] $, we find:
\[ 
  \begin{array}{l}
(P_1-z)^{-1}\circ \chi _0=\chi _1 \circ (P_0-z)^{-1}\chi_0
-\chi _2\circ (P_0-z)^{-1}[P,\chi_1] (P_0-z)^{-1}\chi_0+ \cdots \\
\pm \chi_N (P_0-z)^{-1}[P,\chi_{N-1}](P_0-z)^{-1} [P,\chi_{N-2}]
\cdots  (P_0-z)^{-1}\chi_0\\
\mp (P_1-z)^{-1}[P,\chi_{N}](P_0-z)^{-1} 
\cdots  (P_0-z)^{-1}\chi_0\end{array}
\]

Let us give now  $\chi_0,~\psi \in C_o^\infty (Y)$ with
  disjoints supports.  By 
 choosing the $\chi_l's$ for $l>0$ with supports disjoint from the
 support
of $\psi$, 
 we see, using Equation (\ref{equ:1}),   that,
 for any $N$,
\[
\| \psi (P_1-z)^{-1}\chi_0 \|_{0,2}=O\left(\hbar^N |\Im z|^{-(N+1)}\right)~.
\] 
The standard a priori elliptic estimates
\[ \| u\|_{s+2,\Omega _1} \leq C
\left( \|(P-z) u\|_{s,\Omega _2}+ \| u\|_{s,\Omega _1}\right)  \]
for $z\in K \subset \subset \C$ and
$\Omega _1 \subset \subset \Omega _2  \subset \subset \R^d$, allow to 
prove that, for any $N,s$, there exists $M(N,s)$ so that
\begin{equation} \label{equ:smooth}
\| \psi (P_1-z)^{-1}\chi_0 \|_{s,s+N+2}=O(\hbar^N |\Im z|^{-M(N,s)})
\end{equation}
Let $\chi \in C_o^\infty (Y)$ so that $\chi \equiv 1 $ on the support 
of $\chi_0$. Let us apply a multiplication by $\chi _0$
to the right and to the left to Equation (\ref{equ:1})
and choose $\psi $ with support disjoint of $\chi_0$ so that
$[P,\chi ]\psi =[P,\chi ]$.
Inserting $\psi $ this way in Equation (\ref{equ:1}),
we get, using Equation (\ref{equ:smooth}):
\[ \chi _0 (P_1-z)^{-1}\chi_0- \chi _0 (P_0-z)^{-1}\chi_0=K \]
and, for any $N$, there exists $M(N)$ so that
  $\|K \| _{-N,N}=O(\hbar^N \Im z^{-M(N)})$.
We now apply the formula (known to some people as the
``Helffer-Sj\"ostrand
formula'', proved for example in the book \cite{Di-Sj},
p.  94--95), valid for $f\in C_o^\infty (\R)$
and $\tilde{f} $ an almost holomorphic extension of $f$:
\[ f(P_j)= \frac{1}{\pi}\int _\C \pa _{\bar{z}}\tilde{f}(z) 
(P_j-z)^{-1} dL(z) ~,\] 
where $dL(z)$ is the canonical Lebesgue measure in the complex plane.
From this, we see that 
$f(P_0)-f(P_1)$ is smoothing in $Y$.

\end{document}